# Accurate Evaluation on the Interactions of SARS-CoV-2 with its Receptor ACE2 and Antibodies CR3022/CB6

Hong-ming Ding[*], Yue-wen Yin[1], Song-di Ni[1], Yan-jing Sheng[1], Yu-qiang Ma[2,*]

[1] Center for Soft Condensed Matter Physics and Interdisciplinary Research, School of Physical Science and Technology, Soochow University, Suzhou 215006, China.

[2] National Laboratory of Solid State Microstructures and Department of Physics, Collaborative Innovation Center of Advanced Microstructures, Nanjing University, Nanjing 210093, China.

*Email: dinghm@suda.edu.cn, myqiang@nju.edu.cn

## Abstract

The spread of the coronavirus disease 2019 (COVID-19) caused by severe acute respiratory syndrome coronavirus-2 (SARS-CoV-2) has become a global health crisis. The binding affinity of SARS-CoV-2 (in particular the receptor binding domain, RBD) to its receptor angiotensin converting enzyme 2 (ACE2) and the antibodies is of great importance in understanding the infectivity of COVID-19 and evaluating the candidate therapeutic for COVID-19. In this work, we propose a new method based on molecular mechanics/Poisson-Boltzmann surface area (MM/PBSA) to accurately calculate the free energy of SARS-CoV-2 RBD binding to ACE2 and antibodies. The calculated binding free energy of SARS-CoV-2 RBD to ACE2 is -13.3 kcal/mol, and that of SARS-CoV RBD to ACE2 is -11.4 kcal/mol, which agrees well with experimental result (-11.3 kcal/mol and -10.1 kcal/mol, respectively). Moreover, we take two recently reported antibodies as the example, and calculate the free energy of antibodies binding to SARS-CoV-2 RBD, which is also consistent with the experimental findings. Further, within the framework of the modified MM/PBSA, we determine the key residues and the main driving forces for the SARS-CoV-2 RBD/CB6 interaction by the computational alanine scanning method. The present study offers a computationally efficient and numerically reliable method to evaluate the free energy of SARS-CoV-2 binding to other proteins, which may stimulate the development of the therapeutics against the COVID-19 disease in real applications.

# 1. Introduction

By October 31th 2020, the coronavirus disease 2019 (COVID-19) has infected over 46 million individuals and led the death of more than 1.2 million people in the whole world.[1] The COVID-2019 is mainly caused by the novel severe acute respiratory syndrome coronavirus-2 (SARS-CoV-2),[2-4] which is closely related to several bat coronaviruses and to severe acute respiratory syndrome coronavirus (SARS-CoV).[3-5] Similar to SARS-CoV, the spike glycoprotein on the surface of SARS-CoV-2 plays an important role in the virus entry, which can bind the human angiotensin converting enzyme 2 (ACE2) protein via S1 domain (in particular the receptor binding domain, RBD) and fuse with the cell membrane via S2 domain.[6-9] The similarity between the two spike glycoproteins (i.e., SARS-CoV and SARS-CoV-2) was nearly 80%.[5,10] However, SARS-CoV-2 was believed to transmit from human to human more easily (compared to SARS-CoV), thereby caused much more cases.

More recently, monoclonal antibodies targeting the RBD of the spike glycoprotein are increasingly recognized as a promising method of treating COVID-2019.[11-17] For example, Shi et al.[11] isolated two specific human monoclonal antibodies (i.e., CA1 and CB6) from a convalescent COVID-19 patient, and demonstrated potent SARS-CoV-2-specific neutralization activity *in vitro* against SARS-CoV-2. Wang et al.[12] reported an antibody named 47D11 that binds a conserved epitope on the RBD to cross-neutralize SARS-CoV and SARS-CoV-2. Such targeting behavior (distal from the receptor binding site) was also reported by Wilson and coworkers,[13] where they also determined the crystal structures of the complex of antibody CR3022 with the receptor binding domain (RBD).

Apart from the great progress in the experimental and clinical studies, there have been some simulation works that investigated the interaction between ACE2 and RBD at the molecular level.[18-26] For example, Wang et al.[18] revealed that the binding interface (of ACE2-RBD) consists of a primarily hydrophobic region and a delicate hydrogen-bonding network. Amin et al.[20] compared the binding affinities of the RBD of SARS-CoV and SARS-CoV-2 to ACE2 and found that the binding energies at the interface are a bit higher for SARS-CoV-2 because of enhanced electrostatic interactions. Zou et al.[26] performed computational alanine scanning mutagenesis on the hotspot residues using relative binding free energy calculations, and found that the mutations in SARS-CoV-2 led to a greater binding affinity relative to SARS-CoV. More importantly, Han et al.[22] computationally designed some peptide inhibitors extracted from the sequence of ACE2, which showed a highly specific and stable binding (blocking) to SARS-CoV-2.

Notably, it is still hugely challenging to accurately calculate the free energy of SARS-CoV-2 binding to its receptor ACE2 and the antibodies, which can not only provide insights into the ACE2-RBD interaction (in particular the effect of the residue mutation on the transmitting ability), but also guide the experimental design of man-made monoclonal antibodies. However, in highly charged bio-systems, the accuracy of the free energy by traditional molecular mechanics/Poisson-Boltzmann surface area (MM/PBSA) and molecular mechanics/generalized Born surface area (MM/GBSA), especially the MM/PBSA, could be very poor.[27-28] On the other hand, although the prediction by alchemical free energy (AFE) methods like free energy perturbation (FEP) and thermodynamic integration (TI) is believed to be more accurate, the computational cost of these methods is extremely expensive,[29] and more importantly the convergence of the FEP and TI is also very difficult in charged systems.[30-31] For example, the experimental results all demonstrate that the binding affinity of SARS-CoV-2 to ACE2 was about -13.0~-10.0 kcal/mol,[6,10,32] and was a little stronger than that of SARS-CoV to ACE2 (although the absolute value of the binding affinity was varied among different experiments). However, the calculated binding free energy by different simulation methods was from dozens to hundreds of kcal/mol, [20-25] which is much stronger than that in real experiment. Some simulation studies even indicated that the binding affinity of SARS-CoV-2 to ACE2 was weaker than that of SARS-CoV to ACE2,[24-25] which certainly contradicted with the main experimental findings. Thus, presently there existed a big gap between the prediction by the simulation method and that in real experiments, which urgently requires the improvement of free-energy methods.

In this study, we propose a new MM/PBSA method based on the screening electrostatic energy in molecular mechanics to evaluate the binding free energy of SARS-CoV-2 and other proteins. As we will show below, when combining with the interaction entropy method, the predicted binding free energy of SARS-CoV-2 to ACE2 and SARS-CoV to ACE2 is in good agreement with the experimental result. Moreover, the binding free energy of SARS-CoV-2 to two antibodies (i.e., CR3022 and CB6) with potential clinical use is evaluated based on the proposed method, and also agrees with the experimental result. Further, the computational alanine scanning demonstrates the significant role of R97 in CB6 and R403 in SARS-CoV-2 RBD as well as the hydrophobic contacts in the CB6/SARS-CoV-2 interaction.

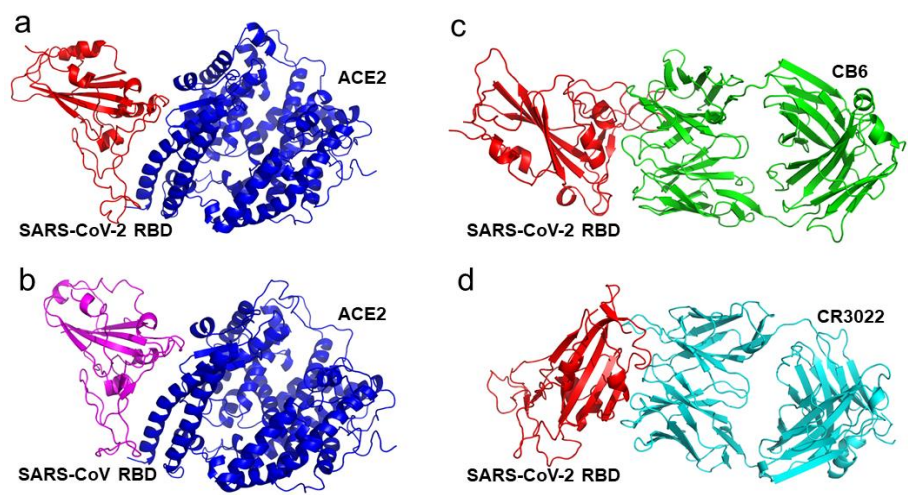

**Figure 1.** Crystal structures of the four complexes used in this work. (a) SARS-CoV-2 (RBD)/ACE2 (PDB ID: 6M0J), (b) SARS-CoV (RBD)/ACE2 (PDB ID: 2AJF), (c) SARS-CoV-2 (RBD)/CB6 (PDB ID: 7C01), and (d) SARS-CoV-2 (RBD)/CR3022 (PDB ID: 6W41).

## 2. Modeling and Methods

As shown in Figure 1, four different protein complexes were considered in this work, namely SARS-CoV-2 (RBD)/ACE2 (PDB ID: 6M0J),[10] SARS-CoV (RBD)/ACE2 (PDB ID: 2AJF),[33] SARS-CoV-2 (RBD)/CB6 (PDB ID: 7C01),[11] and SARS-CoV-2 (RBD)/CR3022 (PDB ID: 6W41).[13] Each complex was solvated in TIP3P water[34] (the minimum distances from the surfaces of the box to the complex atoms were set to 15Å), with NaCl to neutralize the systems at the concentration of 0.15 M. The system was firstly energy-minimized by the steepest descent method until the convergence was reached. Then the system was gradually heated from 0 to 298 K in the NVT ensemble over a period of 500 ps, and then relaxed by 500 ps in the NPT ensemble with 1000 kJ*mol$^{-1}$*nm$^{-2}$ harmonic constraints on the heavy atoms of the proteins, where the temperature was controlled at 298 K by the V-rescale thermostat with a time constant of 0.2 ps and the pressure was kept at 1 atm by the Berendsen barostat with a time constant of 2.0 ps. Finally, 2 ns NPT simulations with weak constrains (100 kJ*mol$^{-1}$*nm$^{-2}$) on the heavy atoms of the proteins were performed for each system, which could prevent the structural drift to the incorrect protein structures.[35-36] All the MD simulations were carried out by using GROMACS 2019.03 package[37] with Amber ff14sb force field.[38] The LINCS constraints were used to all bonds involving hydrogen atoms.[39] The particle mesh Ewald method was used when calculating the long-range electrostatic interactions,[40] and the Lennard-Jones (LJ) interactions were cut off at a distance of 1.0 nm. The periodic boundary conditions were adopted in all three directions.

In standard MM/PBSA, the binding free energy ($\Delta G_{bind}$) between the proteins is calculated as:[41-42]

$$\Delta G_{bind} = G_{com} - (G_{proA} + G_{proB}),$$
$$\Delta G_{bind} = \Delta H - T\Delta S \approx \Delta E_{MM} + \Delta G_{sol} - T\Delta S,$$
$$\Delta E_{MM} = \Delta E_{int} + \Delta E_{ele} + \Delta E_{vdw},$$
$$\Delta G_{sol} = \Delta G_{PB} + \Delta G_{SA},$$

where $G_{com}$ is the free energy of the complex, and $G_{proA}$ is the free energy of one protein and $G_{proB}$ is the free energy of the other protein. $\Delta G_{bind}$ can be usually decomposed into three terms: the gas-phase interaction energy $\Delta E_{MM}$, the desolvation free energy $\Delta G_{sol}$, and the conformational entropy $-T\Delta S$.

$\Delta E_{MM}$ is composed of $\Delta E_{int}$, $\Delta E_{vdw}$, and $\Delta E_{ele}$. Since there were no bonds between the proteins, $\Delta E_{int}$ was zero in this case. $\Delta E_{vdw}$ is the VDW interaction energy between the proteins, and calculated by the Lennard-Jones (12-6) potentials. $\Delta E_{ele}$ is the electrostatic energy between the proteins, and is usually calculated by using the following equation:

$$\Delta E_{ele} = \sum_i^{proA} \sum_j^{proB} \frac{q_i q_j}{4\pi\varepsilon_0 \varepsilon_{in} r_{ij}},$$

where $q_i$, $q_j$ is the charge of the atom $i$, atom $j$ in the two proteins, respectively, $r_{ij}$ is the distance between atom $i$ and atom $j$, $\varepsilon_0$ is the dielectric constant in the vacuum, and $\varepsilon_{in}$ is the relative dielectric constant of the solute (i.e., the protein).

However, since the proteins carry some net charges, there are a lot of counter-ions around them, thus the effective electrostatic interaction between them is greatly changed. To mimic the screening effect, we added an additional exponential damping to the Columbic term according to Debye−Huckel theory,[43-44] i.e.,

$$\Delta E_{ele} = \sum_i^{proA} \sum_j^{proB} \frac{q_i q_j}{4\pi\varepsilon_0 \varepsilon_{in} r_{ij}} e^{-r_{ij}/\lambda_D},$$

where $\lambda_D$ is the Debye length, and can be determined by Debye−Huckel theory:[43, 44]

$$\lambda_D = \sqrt{\varepsilon_0 \varepsilon_r k_B T \Big/ \sum_i c_i e^2 z_i^2},$$

where $\varepsilon_r$ is the relative dielectric constant of the medium/solvent, $c_i$ and $z_i$ is the concentration and the net charge of ion $i$, respectively, the sum is over all ionic species. When the type of the salt is NaCl (0.15 M) and the temperature is 298K, the Debye length can be calculated as $\lambda_D = 8.0$ Å.

$\Delta G_{sol}$ is the sum of $\Delta G_{SA}$ (non-polar contribution) and $\Delta G_{PB}$ (polar contribution). The former one is usually estimated using the solvent accessible surface area (SASA), i.e., $\Delta G_{SA} = \gamma \cdot SASA + b$, where $\gamma$ is the surface tension 0.0227 kJ*mol$^{-1}$*Å$^{-2}$ and $b$ is the constant 3.8493 kJ*mol$^{-1}$. The latter one is calculated by using the Poisson-Boltzmann (PB) model; due to the highly charged property of ACE2, the electrostatic potential was obtained by solving the non-linear Poisson- Boltzmann Equation (nPBE) in this work.[42]

The entropy term is calculated by the interaction entropy (IE) method,[45-46] by which one can calculate the entropic term directly from the molecular dynamics simulations without any extra computational cost. The definition of the interaction entropy is as follows:

$$-T\Delta S = k_B T \ln \langle e^{\Delta\Delta E_{MM}/k_B T} \rangle = k_B T \ln\left(\frac{1}{N}\sum_{i=1}^{N} e^{\Delta\Delta E_{MM}(i)/k_B T}\right),$$

where $\Delta\Delta E_{MM}(i) = \Delta E_{MM}(i) - \langle \Delta E_{MM} \rangle$, and denotes the fluctuation of

protein−protein interaction energy around the average energy, $k_B$ is the Boltzmann constant.

In the binding free energy calculation, 100 frames (with an interval of 10 ps in the last 1 ns) was used for calculating the binding energy in each system via the MM/PBSA method. All the MM/PBSA calculations were performed by using the modified shell script gmx_mmpbsa.[47] 10000 frames (with an interval of 0.1 ps) were used for calculating the entropic term via the interaction entropy (IE) method. Three independent runs were performed for each system to obtained the standard deviation. Moreover, the computational alanine scanning was carried with the simple mutations within the framework of the single-trajectory method.[48-49] In more detail, the mutated trajectory was simply generated from the wild-type trajectory, where the side chain of the residue was truncated and the $C_\beta$ atom and its linking hydrogen atoms were retained. Then a hydrogen atom was added in the same direction as that of the $C_\beta$–X (the truncated atom) bond and the length of the new $C_\beta$–H bond was set to the default value in the force field. The simple mutation is an easy, efficient and fast mutagenesis method to identify the hot spots at the binding interface. Yet one should be careful when the conformational change of proteins is obvious due to the mutation.[50] Additionally, this method cannot be applied when mutating a smaller residue to a larger one.[48]

## 3. Results and Discussion

The free energy of SARS-CoV-2 binding to ACE2 calculated by standard MM/PBSA and screening MM/PBSA is listed in Table S1. Since ACE2 carries a net charge of -28e and SARS-CoV-2 RBD carries a net charge of +2e, the attractive electrostatic energy would be very strong when neglecting the screening effect under the dielectric constant 2.0 (a typical setup for biomolecules), which led to the extremely low binding energy (Figure 2). We further increased the dielectric constant in standard MM/PBSA. Even if the dielectric constant was set as 17.0 (Table S1), the electrostatic energy was greatly reduced but the PB energy also decreased a lot, the predicted binding free energy (~-55.4 kcal/mol) was still much larger than the experimental result (~-11.3 kcal/mol). On the contrary, the binding free energy predicted by the screening MM/PBSA was about -13.3 kcal/mol, which was close to that in the experiment.

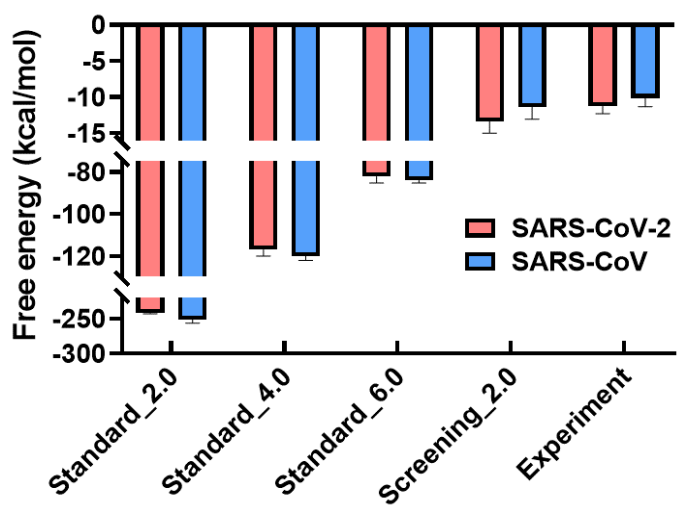

**Figure 2.** The comparison of the free energy among the standard MM/PBSA (using different $\varepsilon_{in}$), the screening MM/PBSA ($\varepsilon_{in}$=2.0), and the experiment result in the cases of SARS-CoV-2 RBD/ACE2 and SARS-CoV RBD/ACE2. The standard deviation for simulation data were obtained by three independent runs, and that for the experimental data were obtained by three experimental works.[6,10,32]

To test the robustness of the binding free energy calculated by the screening MM/PBSA, we systematically investigate the effect of the length of simulation time, the temperature, as well as the water molecules on the results. As shown in Figure S1, the RMSD was nearly the same at different temperature. Although the temperature may also change the relative dielectric constant of the water and the Debye length, the change was very little, thus the binding free energy and the energy terms were nearly the same at 298 K and 310 K. Similarly, the time length had very little impact on the binding free energy (Figure S2). Moreover, we also explicitly considered the associated water molecules at the binding site in MM/PBSA. In this case, although the energy terms changed obviously, namely the electrostatic energy was enhanced and at the same time the PB energy was increased (and balanced out the increase of the electrostatic energy), the total binding energy changed little in the presence of water molecules (Figure S3). This is probably due to the weak association of water molecules with the two proteins at the binding site.

We also calculated the residue contact probability to determine the binding interface (Figure S4) and analyzed the key residues in the SARS-CoV-2-ACE2 interaction via the residue energy decomposition in MM/PBSA. Figure 3 illustrates the residues of top ten binding energies in SARS-CoV-2 RBD and ACE2, respectively. In screening MM/PBSA, most of these residues were at the binding site except few charged residues like K444 in SARS-CoV-2 and E23 in ACE2 (these charged residues were also near the binding site, see Figure 3a-b). Notably, previous studies[10,18] indicated the significant roles of Y449, N487, T500, G502, Y505 in forming hydrogen bonds, and that of F456, Y473, F486, Y489 in forming the hydrophobic pockets, most of which were listed in the top ten residues here. Nevertheless, since the standard MM/PBSA overestimated the attractively electrostatic interaction in the long range, most of the residues were charged residues and the binding energy per residue was extremely strong in SARS-CoV-2 (Figure 3c). Besides, these top ten residues were far away from the binding site (Figure 3d), which was certainly unreasonable. We further analyzed the binding energy inside/outside the binding site with screening MM/PBSA and standard MM/PBSA, respectively. As shown in Figure S5, the standard MM/PBSA showed that the binding site played little role in the binding energy since the electrostatic energy was very small compared to the total electrostatic energy. As a result, the binding energy at the binding site accounted for very small proportion in the total binding energy in standard MM/PBSA. On the contrary, the screening MM/PBSA indicated the significant role of the binding site in the total binding energy, where the electrostatic energy was dominated at the binding site.

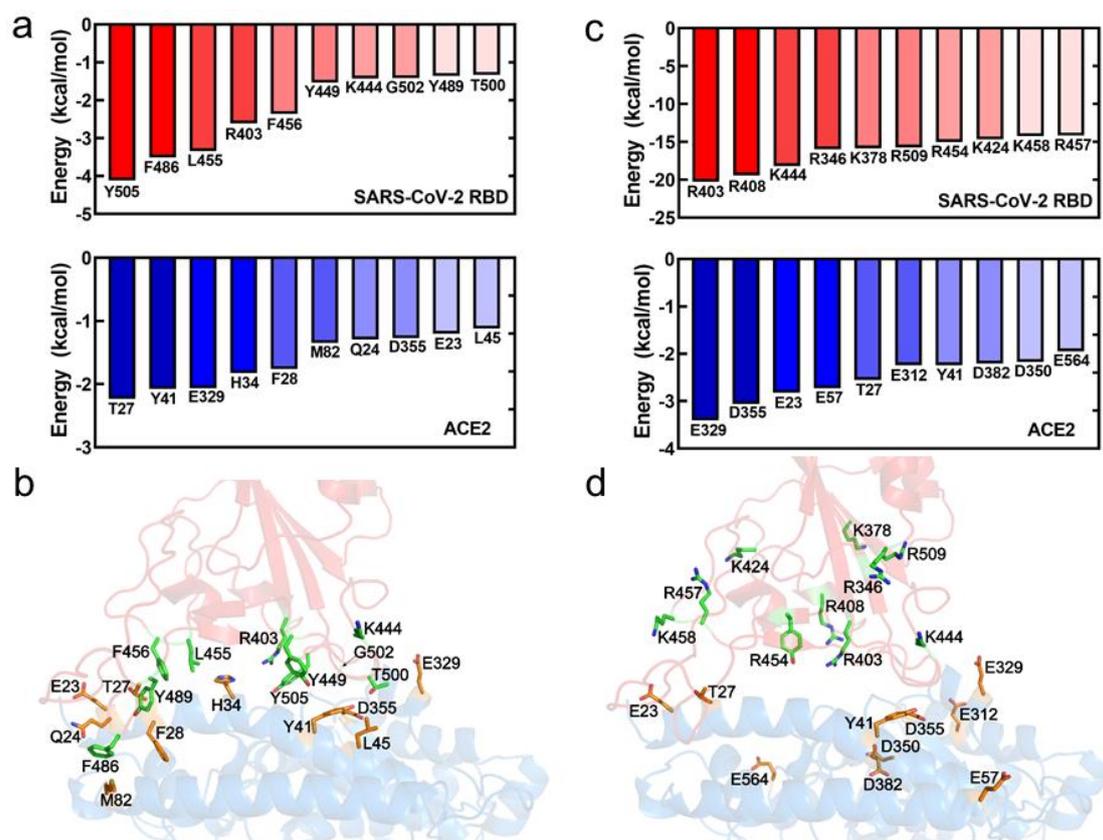

**Figure 3.** Decomposition of the energy per residue in the interaction between the RBD of SARS-CoV-2 and ACE2. (a) The top-ten per-residue binding energy determined by the screening MM/PBSA ($\varepsilon_{in}$=2.0), most of the residues were near the binding site (b); (c) The top-ten per-residue binding energy determined by the standard MM/PBSA ($\varepsilon_{in}$=6.0), most of the residues were far away from the binding site (d).

We then calculated the free energy of SARS-CoV binding to ACE2 by using the MM/PBSA (Table S2). Similar to SARS-CoV-2, the RBD of SARS-CoV also carries a net charge of +2e, thus the free energy predicted by standard MM/PBSA was still very low due to the strong electrostatic energy (even using a large dielectric constant 6.0). Notably, the free energy of SARS-CoV binding to ACE2 was lower than that of SARS-CoV-2 binding to ACE2 in standard MM/PBSA (Figure 2), which was inconsistent with the experiment result. On the contrary, the free energy of SARS-CoV binding to ACE2 was about -11.4 kal/mol in screening MM/PBSA, which was close to the experiment result (-10.1 kal/mol), and was a bit weaker than the predicted free energy of SARS-CoV-2 binding to ACE2 (-13.3 kcal/mol), which was probably due to the weaker electrostatic interactions.[18-20] It is well known that the transmissibility of COVID-2019 is higher than that of SARS-2003.[51-52] Thus, SARS-CoV-2 may have greater ability of entering cell hosts than SARS-CoV. Apart from the unique 'RRAR' furin cleavage site at the S1–S2 boundary,[8,10] the SARS-CoV-2 may have a higher binding affinity (to ACE2), which is supported by our simulation result. Similar to previous case, here we also analyzed the key residues in the SARS-CoV/ACE2 interaction via the residue energy decomposition. Not surprisingly, the standard MM/PBSA showed the wrong trend (i.e., the residues of top ten binding energies were far away from the binding sites and the binding energy per residue was very strong), while the screening MM/PBSA again gave a reasonable result (see Figure S6).

In general, the above results demonstrated that the screening effect between the charged biomolecules should be considered in calculating the binding free energy, particularly when the biomolecules carried a lot of net charges.

The SARS-CoV-2-neutralizing antibodies primarily target the RBD on spike proteins. The binding affinity (of the antibody to RBD) is one of the most important factors in evaluating potential clinic use of an antibody. Here, we first calculated the free energy of SARS-CoV-2 binding to two recently reported antibodies (i.e., CR3022 and CB6) using the screening MM/PBSA. As shown in Figure 4 (see details in Table S3 and Table S4), the free energy of SARS-CoV-2 binding to CR3022 was -11.3 kcal/mol, and that of SARS-CoV-2 binding to CB6 was -17.4 kcal/mol, which consisted with the experimental findings (i.e., -9.5 kcal/mol,[13] -11.7 kal/mol,[11] respectively) in ranking, again indicating the good performance of the screening MM/PBSA. Notably, although the electrostatic energy in the case of CR3022 was stronger than that in the case of CB6, the larger PB energy and the weaker VDW energy led to the weaker binding affinity of CR3022 to SARS-CoV-2.

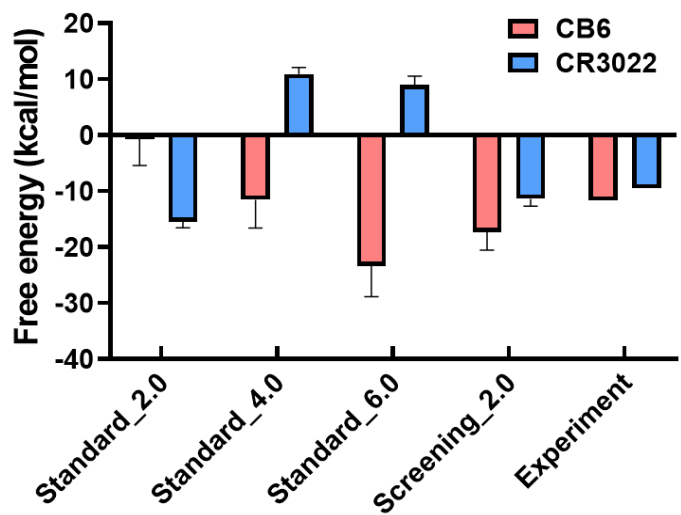

**Figure 4.** The comparison of the free energy among the standard MM/PBSA (using different $\varepsilon_{in}$), the screening MM/PBSA ($\varepsilon_{in}$=2.0), and the experiment result in the cases of SARS-CoV-2 RBD/CB6 and SARS-CoV-2 RBD/CR3022. The standard deviation for the simulation data were obtained by three independent runs. The experimental data for CB6 and CR3022 was taken from ref. [11] and ref. [13], respectively.

We further evaluated the accuracy of standard MM/PBSA in this case. Unfortunately, although the standard MM/PBSA predicted a comparative binding energy (-15.5 kcal/mol for CR3022 and -0.8 kcal/mol for CB6) to that in the experiment (-9.5 kcal/mol and -11.7 kcal/mol), it reported the wrong trend for the two antibodies at $\varepsilon_{in}$ =2.0. When increasing $\varepsilon_{in}$ to 4.0 or 6.0, the standard MM/PBSA predicted the right trend (i.e., $\Delta G_{CB6} < \Delta G_{CR3022}$), but the binding free energy of CR3022 (to RBD) became positive. In general, the performance of standard MM/PBSA was still poor in this case.

Having demonstrated that the screening MM/PBSA could provide relatively accurate binding free energy, we further analyzed the key residues at the binding interface between the antibodies CR3022/CB6 and SARS-CoV-2 and investigated whether the binding affinity of antibodies could be enhanced or weakened with the computational alanine scanning method.

The residue contacting probability of antibodies binding to SARS-CoV-2 was first calculated, where the residues at the interface were illustrated in Figure S7. Obviously, the heavy chain in both antibodies was of the great importance since there are 23 (17) residues of the heavy chain and only 6 (9) residues of the light chain in the binding sites in the case of CB6 (CR3022). The contact surface area of the heavy chains with SARS-CoV-2 was also much larger than that of the light chains (Figure 5a-b). More importantly, Figure S8 shows that 63% of the epitopes (i.e., SARS-CoV-2/ACE2 binding sites) was occupied by CB6, which could greatly decrease the probability of SARS-CoV-2 binding to ACE2. On the contrary, no epitopes in SARS-CoV-2 were occupied by CR3022. Nevertheless, the epitopes of RBD that CR3022 targets are highly conserved between the SARS-CoV-2 and SARS-CoV, thus CR3022 or some like antibodies may have the potential of serving as a universal antibody for coronaviruses.[12-13]

To provide more detailed insight into the interaction of the antibody with the RBD of SARS-CoV-2, we took the CB6 as an example and identified the hot-spot residues with the computational alanine scanning method (see Table S5). Figure 5c shows that there were four hot spots ($\Delta\Delta G > 2.0$ kcal/mol) and seven warm spots (1.0 kcal/mol < $\Delta\Delta G$ < 2.0 kcal/mol) in CB6 at the binding interface. Most of the hot (4/4) and warm spots (6/7) were in the heavy chain, again indicating the important role of the heavy chain in neutralizing the antigen in CB6. Moreover, we also investigated the effect of the residue mutation of epitopes in RBD on the binding affinity since SARS-CoV-2 is believed to mutate more quickly than some other viruses. It was found that most of the hot and warm spots were dominated by the VDW interaction (Table S5), and there was a large hydrophobic domain in the CB6-RBD interaction (see the blue circle in Figure 5d). Moreover, the hydrogen bonds also played an important role in the interaction. For example, the highest value of $\Delta\Delta G$ in CB6 was R97 in the heavy chain, which formed two hydrogen bonds with N487 in SARS-CoV-2 (Figure 5d). Moreover, the highest value of $\Delta\Delta G$ in SARS-CoV-2 was R403, which also formed two hydrogen bonds with Y92 in the light chain of CB6.

We further made an attempt to evaluate whether and how the mutations (to the other residues) in the antibody could improve the binding affinity since the antibody could mutate quickly in B cells.[53] As discussed above, the interaction of R403, R408 in RBD with Y92, P95 in the light chain of CB6 and the hydrophobic contacts between the RBD and the heavy chain in CB6 were both important. Considering that R403 and R408 are positively charged, it may be further enhanced when Y92 and P95 mutated to negatively charged residues. Moreover, increasing the aromatic residues at the hydrophobic contact region may also have the probability of increasing the binding affinity. The above inference was indeed verified as shown in Figure S9.

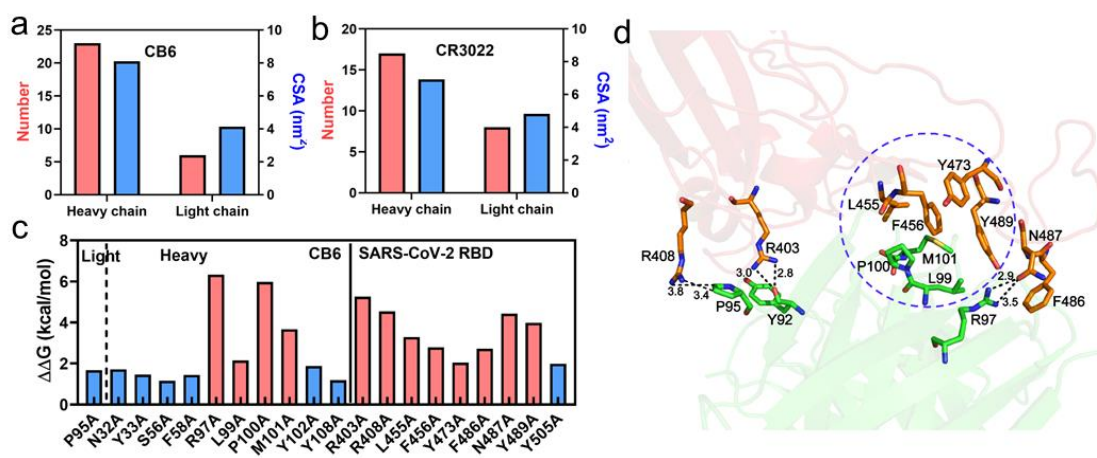

**Figure 5.** Interaction between the antibodies and SARS-CoV-2 RBD. The number of contacting residues and the contact surface area of the light chain and heavy chain in the case of SARS-CoV-2/CB6 (a) and SARS-CoV-2/CR3022 (b). (c) Hot spots (red bars) and warm spots (blue bars) in the SARS-CoV-2/CB6 interaction calculated by the computational alanine scanning method within the framework of screening MM/PBSA. (d) Snapshot of key residues engaging hydrogen bonds and hydrophobic interactions at the SARS-CoV-2/CB6 interface. The hydrogen bonds are highlighted as black dash lines and the hydrophobic contact region is highlighted as a blue dash circle.

## 4. Conclusions

In summary, we have proposed a new method (i.e., screening MM/PBSA) to evaluate the binding free energy between SARS-CoV-2 RBD, ACE2 as well as two antibodies. It was found that the free energies of SARS-CoV-2 RBD to ACE2 and SARS-CoV RBD to ACE2 calculated by standard MM/PBSA were much lower than the experimental result and even showed the wrong trend. By introducing the screening electrostatic energy in MM/PBSA, the calculating free energies were in good agreement with the experimental result. Moreover, we evaluated the performance of the screening MM/PBSA on the binding affinity of antibodies to SARS-CoV-2 RBD, and found that the predicted binding free energy also agreed with the experimental findings (the performance of standard MM/PBSA was still poor in this case). In general, the screening MM/PBSA can give a more reliable prediction of the binding free energy of SARS-CoV-2 RBD-involved interactions, and should have great potential in evaluating the protein-protein interaction in highly charged bio-systems.

**Conflict of interest**

The authors declare no competing financial interest.

**Acknowledgments**

This work is supported by the National Natural Science Foundation of China (Nos. 11874045 and 11774147).